\documentclass[aps,prl,twocolumn,superscriptaddress,showpacs]{revtex4}
\usepackage{graphicx}
\usepackage{dcolumn}
\usepackage{bm}
\usepackage{color}
\usepackage[colorlinks]{hyperref}
\input{epsf}

\begin{document}


\title{Radiation induced zero-resistance states in GaAs/AlGaAs
heterostructures: Voltage-current characteristics and intensity
dependence at the resistance minima }

\author{R. G. Mani}
\email{mani@deas.harvard.edu}
\affiliation {Harvard University,
Gordon McKay Laboratory of Applied Science, 9 Oxford Street,
Cambridge, MA 02138, USA}

\author{V. Narayanamurti}
\affiliation {Harvard University, Gordon McKay Laboratory of
Applied Science, 9 Oxford Street, Cambridge, MA 02138, USA}

\author{K. von Klitzing}
\affiliation {Max-Planck-Institut f\"{u}r Festk\"{o}rperforschung,
Heisenbergstrasse 1, 70569 Stuttgart, Germany}

\author{J. H. Smet}
\affiliation{Max-Planck-Institut f\"{u}r Festk\"{o}rperforschung,
Heisenbergstrasse 1, 70569 Stuttgart, Germany}

\author{W. B. Johnson}
\affiliation {Laboratory for Physical Sciences, University of
Maryland, College Park, MD 20740, USA}

\author{V. Umansky}
\affiliation{Braun Center for Submicron Research, Weizmann
Institute, Rehovot 76100, Israel}

%
%
%
%
\date{\today}
\begin{abstract}
High mobility two-dimensional electron systems exhibit vanishing
resistance over broad magnetic field intervals upon excitation
with microwaves, with a characteristic reduction of the resistance
with increasing radiation intensity at the resistance minima.
Here, we report experimental results examining the voltage -
current characteristics, and the resistance at the minima versus
the microwave power. The findings indicate that a non-linear $V-I$
curve in the absence of microwave excitation becomes linearized
under irradiation, unlike expectations, and they suggest a
similarity between the roles of the radiation intensity and the
inverse temperature.
\end{abstract}
%
%
\maketitle
\section{introduction}

Vanishing resistance induced by electromagnetic wave excitation in
the ultra high-mobility 2-dimensional electron system (2DES), at
low temperatures ($T$) and modest magnetic fields ($B$), has
suggested the possibility of novel non-equilibrium zero-resistance
states (ZRS) in the 2DES.\cite{1} Remarkably, in this effect,
vanishing resistance does not produce plateaus in the Hall
resistance, although the diagonal resistance exhibits activated
transport and zero-resistance states, similar to quantum Hall
effects (QHE).\cite{1,2,3,4,5}

A surge in theory has already produced a physical framework for
viewing this extraordinary phenomenon, with testable predictions
for further
experiment.\cite{6,7,8,9,10,11,12,13,14,15,16,17,18,19,20,21,22,23,24,25,26,27,28,29,30}
A class of hypothesis attributes radiation induced resistance
oscillations to $B$-dependent enhancement/reduction of the
current.\cite{6,8,10,12} An oscillatory density-of-states, an
electric field, and impurity scattering appear to be the key
ingredients in theory, which indicates negative resistivity for
sufficiently large radiation intensities.\cite{6,8,12} The
modeling by Durst and co-workers,\cite{8} which is similar to the
work by Ryzhii,\cite{6} reproduced the period and, approximately,
the phase reported in Ref.1,\cite{1} although there was variance
between theory and experiment so far as the realization of
negative resistivity was concerned.\cite{6,8} A possible physical
instability for a negative resistivity state, meanwhile, led to
the conjecture by Andreev \textit{et al.} of a current dependent
resistivity, and the formation of current domains, along with a
scenario for realizing zero-resistance in measurement.\cite{9}
These works taken together seem to provide a path for realizing
radiation-induced oscillatory magnetoresistance and
zero-resistance.\cite{6,8,9} Yet, there do occur real differences
between the theoretical modeling and experiment. For example, the
type of scattering potential invoked in the Durst \textit{et al.}
theory differs from the experimental situation, where measurements
involve low-disorder, extremely high mobility specimens including
mostly small angle scattering.\cite{1,8} In addition, the current
domain picture of Andreev \textit{et al.} includes boundary
conditions,\cite{9} which appear difficult to realize in the Hall
bar type specimen.\cite{31}

A complementary approach,\cite{12} which modeled the specimen as a
tunnel junction, also realized magnetoresistance oscillations with
a period and phase that were approximately consistent with Ref.
1.\cite{1,12} This theory indicated, in addition, an N-shape
current-voltage ($I-V$) characteristic in a regime of
radiation-induced negative conductivity, analogous to Andreev
\textit{et al}.\cite{9,12} Subsequent theoretical work by
Bergeret, Huckestein, and Volkov provided clarification and
proposed, in the strong $B$-field limit, an S-shape $I-V$
characteristic for Hall bar devices and a N-shape $I-V$
characteristic for the Corbino device,\cite{17} the geometry
dependence following from the boundary conditions in the 2DES.
These works impressed the idea that $I-V$ characteristics of the
ZRS, which appear calculable from theory, might serve as an
incisive and confirming probe of the underlying physics.

In short, the possible transformation of a predicted negative
resistivity/conductivity state, into the vanishing resistance
state reported by experiment, in a strong B-field limit where
vanishing resistance and vanishing conductance are equivalent, and
the associated $I-V$ characteristics, appear to be issues in the
theoretical discussion.\cite{1,2,6,8,9,10,12,14,16,17}

In comparison, experiment indicates a ZRS which is approached
exponentially vs. $T$, following $R_{xx} \approx \exp
(-\Delta/k_{B}T)$, where $\Delta$ is an activation
energy.\cite{1,2} By invoking an analogy to QHE, where large
amplitude (Shubnikov-de Haas) resistance oscillations show similar
ZRS while exhibiting activated transport,\cite{4} such transport
behavior has been cited as evidence for a radiation-induced gap in
the electronic spectrum.\cite{1}

Shi and Xie have argued that a negative conductance instability
could induce a state transition and a new electronic phase, which
includes a physical rearrangement of the system.\cite{12,16} It
appears that this possibility could also be addressed by comparing
experiment with model
predictions.\cite{5,6,7,8,9,10,11,12,13,14,15,16,17,18,19,20,21,22,23,24,25,26,27,28,29,30}
Thus, we examine the radiation induced vanishing resistance states
and associated voltage-current ($V-I$) characteristics in ultra
high mobility GaAs/AlGaAs heterostructures. In a Hall geometry, we
demonstrate that the non-linear $V-I$ curve at $B = (4/5) B_{f}$
in the absence of radiation becomes linearized as $R_{xx}$
$\longrightarrow$ 0 under the influence of radiation, to
relatively high currents. The power variation of $R_{xx}$ at the
minima is also reported and these data suggest that a variable
such as $log(P)$, where $P$ is the radiation power, might play a
role that is approximately analogous to the inverse temperature in
activation studies. The observed differences between experiment
and available model predictions seem to suggest that additional
physics might be involved in the ZRS that is observed in the
irradiated-2DES.
\section{experiment}
Experiments were carried out on Hall bars, square shaped devices,
and Corbino rings, fabricated from GaAs/AlGaAs heterostructures.
After a brief illumination by a red LED, the best material was
typically characterized by an electron density, $n(4.2$ K)
$\approx 3 \times 10^{11}$ cm$^{-2}$, and an electron mobility
$\mu(1.5$ K) upto  $1.5 \times 10^{7}$ cm$^{2}/$ V s. Low
frequency lock-in based electrical measurements were carried out
with the sample mounted inside a waveguide and immersed in pumped
liquid He-3, over the $T$-range $0.4 \leq T \leq 3$ K. Lock-in
based $V - I$ measurements were carried out by quasi-statically
ramping the excitation voltage in the ac constant-current circuit.
Electromagnetic (EM) waves in the microwave part of the spectrum,
$27 \leq f \leq 170$ GHz, were generated using various tunable
sources. The radiation intensity was set at the source and
subsequently reduced using variable attenuators for the reported
experiments at $119$ GHz, where the power in the vicinity of the
sample is estimated to be less than $1$ mW. Measurements at $50$
GHz were carried out using a signal generator with calibrated
output power.\cite{32} Thus, the source output power at $50$ GHz
is given in absolute units.
\begin{figure}
\begin{center}
\includegraphics*[scale = 0.25,angle=0,keepaspectratio=true,width=3.0in]{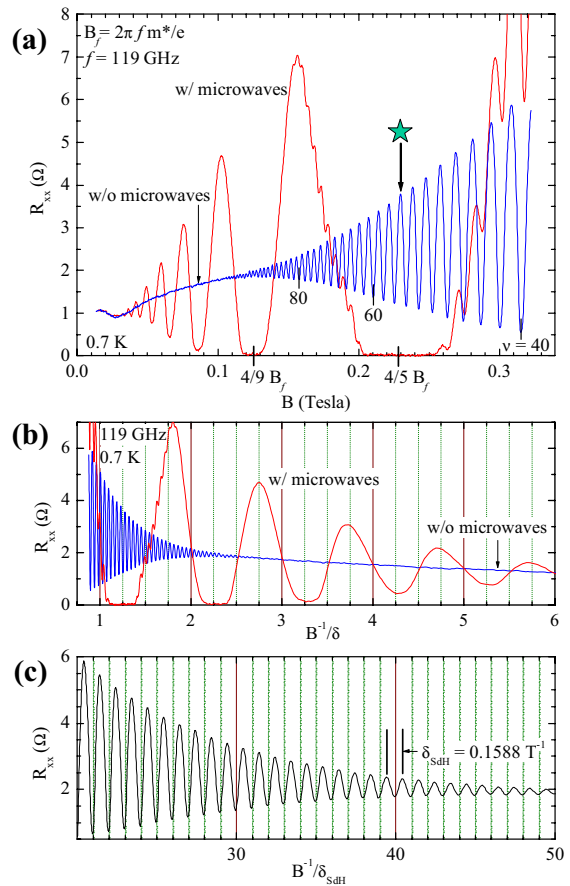}
\caption{(Color online) (a) The magnetoresistance $R_{xx}$ in a
GaAs/AlGaAs heterostructure device with (w/) and without (w/o)
microwave excitation at 119 GHz. Without radiation, Shubnikov-de
Haas (SdH) oscillations are observable in the resistance about
$(4/5) B_{f}$. Radiation reduces the resistance $R_{xx}$ and
eliminates the SdH oscillations in the vicinity of $(4/5) B_{f}$
and $(4/9) B_{f}$. The star marks the B-field at which $V_{xx}$
vs. $I$ measurements are reported (see Fig. 2). (b) A plot of the
resistance data of (a) versus the normalized inverse magnetic
field demonstrates periodicity, with period $\delta$, of the
radiation induced oscillations in $B^{-1}/\delta$. The curves
with- and without- radiation intersect in the vicinity of integral
and half-integral values of the abscissa. (c) A plot of the dark
(w/o radiation) $R_{xx}$ data of (a) versus $B^{-1}/\delta_{SdH}$
helps to determine the phase of the Shubnikov-de Haas
oscillations. Here, $\delta_{SdH}$ is the period of the SdH
oscillations. Note that SdH resistance minima occur at integral
values of $B^{-1}/\delta_{SdH}$, suggesting that $R_{xx}$ is
proportional to $-cos(2\pi F_{SdH}/B)$, where $F_{SdH} =
\delta_{SdH}^{-1}$.}
\end{center}
\end{figure}
\section{results}
Fig. 1 shows the magnetoresistance measured with (w/) and without
(w/o) electromagnetic wave excitation for $B \leq 0.34$ Tesla.
This specimen satisfies the strong field condition,
$\omega_{c}\tau> 1$, for $B > 1 $ mTesla. Here $\omega_{c}$ is the
cyclotron frequency and $\tau$ is the transport relaxation time.
Fig. 1(a) indicates SdH oscillations, which are visible down to $B
\approx 0.1$ Tesla at $0.7$ K, in the absence of radiation. The
application of radiation induces oscillations,\cite{33,34,35,36}
and reduces the resistance over finite $B$-intervals in the
vicinity of $B = [4/(4j+1)] B_{f}$, where $B_{f} = 2\pi f
m^{*}/e$,\cite{1,33,36} $m^{*}$ is the effective mass,\cite{37}
$e$ is the charge of the electron, and $j$ = 1,2,3,... Indeed, ZRS
are visible over broad $B$-intervals in the vicinity of $(4/5)
B_{f}$ and $(4/9) B_{f}$, as the SdH oscillations are suppressed
about the ZRS by the radiation.\cite{1,38,39,40,41,42}

In Fig. 1(b), the data of Fig. 1(a) have been plotted as a
function of the normalized inverse magnetic field, where the
normalization factor, $\delta$, is the period in $B^{-1}$ of the
radiation-induced resistance oscillations.\cite{1} Fig. 1(b)
indicates that the photoexcited (w/radiation) data cross the dark
(w/o radiation) data at integral and half-integral values of the
$B^{-1}/\delta$, when $B^{-1}/\delta \geq 2$ (see also Fig. 3(a)
of Ref. 1). At these $B^{-1}/\delta$, the photon energy $hf$ spans
an integral, $j$, or half integral, $j + 1/2$, cyclotron energies.
The crossing feature in the vicinity of integral $B^{-1}/\delta$
appears to be in agreement with the theoretical
prediction.\cite{8,12} Shi and Xie have suggested that, when $2\pi
f /\omega_{c} = j$, the conductivity in the presence of radiation,
$\sigma$, equals the conductivity in the absence of radiation,
$\sigma_{dark}$, i.e., $\sigma = \sigma_{dark}$.\cite{12}
Remarkably, the data suggest the same behavior at $j + 1/2$.  The
data of Fig. 1(b) also show that resistance minima occur about $j
+ 1/4$, while higher order resistance maxima occur about $j +
3/4$. Experiment suggests that the resistance maxima generally
obey this rule for integral $j$, excepting $j = 0$, where phase
distortion associated with the last peak seems to shift it from
the $2\pi f /\omega_{c} = 3/4$ to approximately $0.85 (\pm 0.03)$
(see Fig. 3(a)).

In order to compare the relative phases of radiation-induced
resistance oscillations and the SdH effect, SdH oscillations have
been shown in a normalized $B^{-1}$ plot in Fig. 1(c), where the
normalization factor $\delta_{SdH}$ is the SdH period in $B^{-1}$.
In Fig. 1(c), the resistance minima of SdH oscillations appear at
integral values ($j = 1,2,3$...) of the abscissa
$B^{-1}/\delta_{SdH}$,\cite{43} implying oscillations of the form
$R_{xx}\approx -\cos(2\pi F_{SdH}/B)$, where $F_{SdH} =
\delta_{SdH}^{-1}$. Thus, there appears to be a phase difference
 between the observed oscillations $(-\cos(2\pi F_{SdH}/B))$ and
the assumed form for the density-of-states (DOS) $(+\cos(2\pi
\varepsilon/\omega_{c}))$ of Ref. 8, which is attributed here to
the suppression of the zero-point energy shift,
$\hbar\omega_{c}/2$, by theory.\cite{8} If this physically
manifested constant energy shift $\hbar\omega_{c}/2$ is included
in $\varepsilon$, then the DOS at low-$B$ looks like
$\eta(\varepsilon) = \eta_{0} - \eta_{1}(\cos(2
\pi\varepsilon/\hbar \omega_{c}))$, with $\eta_{1}$ positive.
Unlike SdH oscillations, the radiation induced oscillations
exhibit minima for $B^{-1}/\delta$ about $j + 1/4$ (see Fig.
1(b)).

\begin{figure}
\begin{center}
\includegraphics*[scale = 0.25,angle=0,keepaspectratio=true,width=3.25in]{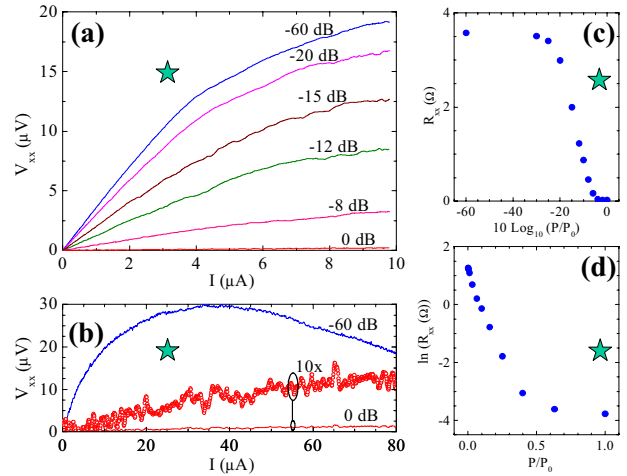}
\caption{(Color online) (a) The diagonal voltage $(V_{xx})$ is
shown as a function of the current $(I)$ for a number of radiation
intensities. The measurements were carried out at the B-field that
is identified by a star in Fig. 1(a). At $(-60$ dB), there is a
sub-linear increase of $V_{xx}$ vs. $I$ above $4 \mu$A. Increasing
the radiation intensity, i.e., tuning the attenuation factor
towards $0$ dB, reduces the slope and linearizes the $V_{xx}$ vs.
$I$ curve. (b) The diagonal voltage $V_{xx}$ is shown as a
function of $I$ to $I = 80 \mu$A, with $(0$ dB) and without $(-60$
dB) microwave excitation. Within experimental resolution, the
$V_{xx}$ vs. $I$ curve with excitation appears mostly linear to
$80 \mu$A. (c) $R_{xx}$ evaluated from the initial slope of the
$V_{xx}$ vs. $I$ curves of (a) are plotted as a function of the
power attenuation factor. Here, $P$ is the attenuated radiation
intensity, and $P_{0}$ is the source intensity. (d) The natural
logarithm of $R_{xx}$ is plotted vs. $P/P_{0}$. } \label{2}
\end{center}
\end{figure}
Results of $V-I$ studies in the vicinity of the $(4/5) B_{f}$
resistance minimum, at the $B$-value marked by the star in Fig.
1(a), are summarized in Fig. 2. Fig. 2(a) shows $V_{xx}$ vs. $I$
to $I = 10$ $\mu$A for several power attenuation factors. In the
absence of radiation ($-60$ dB), $V_{xx}$ vs. $I$ is initially
ohmic at low currents ($\leq 3$ $\mu$A), before non-linearity,
possibly due to heating, becomes apparent above $4$ $\mu$A. The
application of radiation leads to a progressive reduction in the
initial slope of the $V_{xx}$ vs. $I$ curves, signifying a
decrease in $R_{xx}$ in the vicinity of $(4/5) B_{f}$ under
photoexcitation. The data suggest, in addition, that the onset of
non-linearity shifts to higher current with increasing radiation
intensity. This feature implies that the initial linear region
observed in the absence of radiation is extended to higher
currents under the influence of radiation. If the observed
non-linearity originates from heating, then its shift to higher
currents under radiation implies also that heating effects are
reduced under photoexcitation, over the resistance minima.

Experimental results to higher currents shown in Fig. 2(b) confirm
the features observed in Fig. 2(a). Here, the data indicate a
negative differential resistance $r_{xx} = dV_{xx}/dI < 0$ above
$I = 40$ $\mu$A in the absence of radiation ($-60$ dB), while
$r_{xx} > 0$ with microwave excitation ($0$ dB) for the entire
range of currents. Notably, non-linearity in the $V-I$ curve under
radiation, if any, appears only above $50$ $\mu$A, although the
Hall voltage $V_{xy} (10$ $\mu$A) $= 4.69$ mV exceeds the Landau
level spacing $\hbar\omega_{c}/e \approx 0.39$ mV by nearly an
order of magnitude even at $I = 10$ $\mu$A, i.e., $V_{xy} (10$
$\mu$A) $>> \hbar\omega_{c}/e$.

An investigation of the resistance in a Corbino geometry specimen
at $B = (4/5) B_{f}$ and $B = (4/9) B_{f}$ under microwave
excitation also showed current-independent characteristics over
the investigated range, similar to what has been shown for a Hall
geometry in Fig. 3 of Ref. 1. In the Corbino configuration, a
maximum in the resistance (or a minimum in the conductance) was
observed at $B = (4/5) B_{f}$ and $B = (4/9) B_{f}$, supplementing
the usual result for the Hall configuration. The difference was
attributed to the well-known feature of transport, that the
Corbino resistance $R_{C} \approx \sigma_{xx}^{-1}$ while the
diagonal resistance in a Hall geometry $R_{xx} \approx
\sigma_{xx}/\sigma_{xy}^{2}$ under the $\omega_{c}\tau > 1$
condition. Here, $\sigma_{xx}$ and $\sigma_{xy}$ are the diagonal
and off-diagonal components of the conductivity tensor.

The observed change in the $V-I$ characteristics with the
radiation intensity in Fig. 2(a) appears, as mentioned before,
closely correlated to the radiation-induced modification of the
diagonal resistance at the resistance minima. Experiments show
that, at a constant temperature, the resistance minima become
deeper with increasing radiation intensity, until the onset of
"breakdown."\cite{1,40} On the other hand, at constant radiation
intensity, the resistance shows activated transport
characteristics as a function of the temperature.\cite{1,2} Thus,
intuitively, it appears that reducing the temperature and
increasing the radiation intensity play a similar role in the
phenomenology. But the connections are not well understood and,
therefore, the effect of radiation at the deepest resistance
minima is further examined here. The effect of radiation intensity
at the higher order $R_{xx}$ minima have been reported
elsewhere.\cite{38}

\begin{figure}
\begin{center}
\includegraphics*[scale = 0.25,angle=0,keepaspectratio=true,width=3.0in]{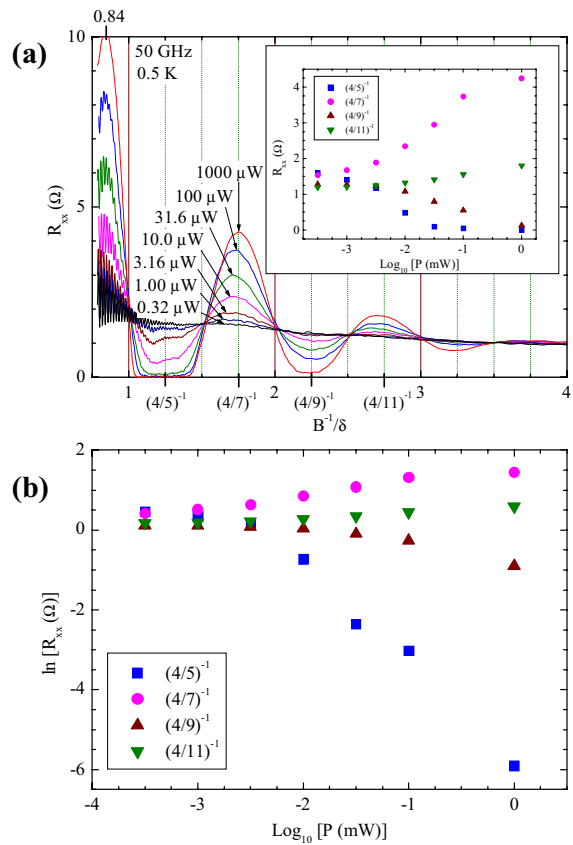}

\caption{(Color online) (a) Radiation induced resistance
oscillations at $f = 50$ GHz are exhibited for a number of source
intensities, in units of $\mu$W. The inset shows the extrema
resistance at $(4/5)^{-1}, (4/7)^{-1}, (4/9)^{-1},  (4/11)^{-1}$
vs. the logarithm of the power. (b) The logarithm of the extrema
resistance vs. the logarithm of the power. A comparison  with
typical activation plots (e.g. see Fig. 3(d) of ref. 1) suggests
that, on the resistance minima, $\log(P)$ might be analogous to
$T^{-1}$. }
\label{4}
\end{center}
\end{figure}
Thus, $R_{xx}$ evaluated from the initial slope of the $V-I$ data
of Fig. 2(a) have been shown in Fig. 2 (c) as a function of $10
Log_{10} (P/P_{0})$, where $P_{0}$ is the (constant) output power
of the radiation source, and $P$ is the attenuated power. This
data plot shows a constant resistance between $-60$ dB and $-30$
dB, followed by a rapid decrease at higher power levels ($> -30$
dB); the resistance becomes small on the exhibited scale above
$-5$ dB. Fig. 2(d) shows an alternate plot with the natural
logarithm of the resistance on the ordinate vs.  the normalized
power on a linear scale along the abscissa. This figure suggests
that, for $P/P_{0} < 0.3$, $R_{xx} \approx \exp [-a(P/P_{0})]$,
with $a \approx$ 12.

Further measurements addressing this issue are illustrated in Fig.
3. Here, vanishing resistance is observable with a source power of
$100$ $\mu W$ about $B^{-1}/\delta = (4/5)^{-1}$. Other features
such as resistance minima near $[4/(4j +1)]^{-1}$, resistance
maxima near $[4/(4j + 3)]^{-1}$ for $j > 1$, nodes in the
resistance in the vicinity of $j$ and $(j + 1/2)$, and the
intersection of the data curves obtained at different powers in
the vicinity of the nodes, are consistent with the results shown
for $f = 119$ GHz in Fig. 1(b). Note also that the first
resistance maximum occurs in the vicinity of $B^{-1}/\delta
\approx 0.84$. That is, this resistance maximum occurs slightly
below $B^{-1}/\delta = 1$, implying a magnetic field for this
feature that is slightly above $B_{f}$. The power dependence of
the resistance at the $(4/5)^{-1}$ and $(4/9)^{-1}$ minima, and
the $(4/7)^{-1}$ and $(4/11)^{-1}$ maxima are shown in the inset
of Fig. 3(a) and in Fig. 3(b). In Fig. 3(b), the ordinate shows
$\ln (R_{xx})$, while the abscissa shows $Log_{10} (P)$.
Remarkably, the plot shown in Fig. 4(b) looks very similar to the
activation plots,\cite{1,2} which suggests that the logarithm of
the radiation power might play a role in power dependent studies
that is analogous to the inverse temperature in activation
studies, at least at the $(4/5)^{-1}$ and $(4/9)^{-1}$ minima.

\section{discussion}

The voltage-current characteristics in the regime of the radiation
induced zero-resistance states could serve to provide valuable
understanding into the physical mechanism underlying the
phenomenon since such characteristics can be measured through
experiment and, at the same time, they appear calculable from
existing theory,\cite{12,14,17} which facilitates a comparison
between experiment and theory.

At the present, it is believed that there ought to be either a
threshold electric field for the onset of current in a Corbino
geometry or a threshold current density for the onset of
dissipation in a Hall geometry.\cite{17} Thus, the naive
expectation has been that a linear voltage-current characteristic
in the absence of radiation should be transformed into a
non-linear characteristic under the influence of radiation.

Theoretical expectations for a non-linear voltage-current
characteristic under photoexcitation are motivated as follows: As
the amplitude of the radiation induced resistivity/conductivity
oscillations increases with increasing radiation intensity, the
oscillatory minima approach and then, in principle, cross over
into a negative conductivity/resistivity regime.\cite{6,8,10,12}
However, the physical instability of negative
resistivity/conductivity prevents the actual observation of
negative resistance/conductance. It leads instead to a locking of
the measured resistance/conductance at
zero-resistance/conductance, along with a stratification of
current into two oppositely directed current domains.\cite{9} As
these current domains are thought to be characterized by an
unspecified current density $j_{0}$, the domains are expected to
rearrange themselves and carry an applied current in a Hall device
without dissipation, so long as the current density associated
with the applied current does not exceed approximately $j_{0}$/2.
Thus, vanishing $V_{xx}$ should be observed below some critical
applied current in a Hall geometry. Above the critical current,
however, there should then be a tendency to destroy the
zero-resistance state as the domains are unable to accommodate the
applied current, and a non-linear voltage-current characteristic
is supposed to reveal such a breakdown of the radiation induced
zero-resistance states with increasing current.

The experimental results for the Hall geometry shown in Fig. 2
appear, however, quite unlike expectations: The voltage - current
characteristic is non-linear in the absence of radiation and it
becomes linearized under the application of radiation. Here, an
unexpected result seems to be the non-linearity and the negative
differential resistance in the absence of radiation in the high
mobility specimen, and the suppression of this non-linearity by
the radiation.

An important supplementary feature of experiment is that, at the
oscillatory resistance minima, there is an activated-type
temperature dependence of the resistance.\cite{1,2} The results
reported here seem to suggest that, in addition to the activated
temperature dependence at the minima, the voltage-current
characteristics constitute yet another thought provoking problem
in this field. The reported voltage-current characteristics, when
coupled with the activated temperature resistance at the minima,
seem to leave open the possibility  that the observed
zero-resistance states might be a consequence of a vanishing bulk
diagonal resistivity, similar to the quantum Hall situation. That
is, it could be that the diagonal resistance saturates at
zero-resistance in activated fashion under photo-excitation
because the diagonal resistivity approaches zero-resistivity in a
similar way. Further experiments seem necessary, however, to
clarify this point.

\section{Summary}

In summary, we have examined the voltage-current characteristics
and the role of the microwave power in the novel radiation induced
zero-resistance states in the high mobility 2DES. The $V-I$
measurement suggest a linearization of the device characteristics
with the application of radiation, while the power dependence at
the minima suggest that the logarithm of the radiation intensity
might play a role that is similar to the inverse temperature in
activation studies. To our knowledge, these features have not been
predicted by theory and, therefore, they seem to provide further
motivation for the study- of this remarkable new effect.\cite{3}
\vspace{0cm}

\end{document}